\documentclass[12pt]{iopart}

\usepackage{cite}
\usepackage{braket}
\usepackage{bm}
\usepackage{graphicx} 
\usepackage{booktabs}
\usepackage{color}

\begin{document}

\title[Relative descriptors for quantum agents]{Relative descriptors for quantum agents}

\author{David Möckli \& Lorenzo Spies Perraro}

\address{Instituto de Física, Universidade Federal do Rio Grande do Sul, Av. Bento Gonçalves 9500, Porto Alegre, 91501-970, RS, Brazil}
\ead{mockli@ufrgs.br}
\vspace{10pt}
\begin{indented}
\item[]October 2024
\end{indented}

\begin{abstract}
We use the Heisenberg picture of Everettian quantum mechanics to work out the relative descriptors for the Frauchiger-Renner thought experiment. While Everettian mechanics reveals no paradox within the thought experiment, the decoherence-free setup offers an instructive branching tree. Traditionally, branching structures derived from the Schrödinger picture suggest that foliations are always sharply defined. However, the Heisenberg picture demonstrates that the tree contains regions with inherently non-sharp foliations, leading to the conclusion that, in Everettian mechanics, quantum agents possess a history with a non-sharp memory registry.
\end{abstract}

\vspace{2pc}

\section{Introduction}\label{sec:introduction}

\textit{Everettian quantum theory} is a realist interpretation of unitary quantum mechanics that seeks to remove the arbitrariness and anthropocentric biases present in standard interpretations.
Measurement outcomes are explained by the relative state construction, which is most commonly understood in the Schrödinger picture. The relative state construction, introduced by Everett almost 80 years ago \cite{Everett1957}, has significantly evolved over time. This evolution is summarised in references \cite{Dewitt_Bryce_2016, saunders2010, Wallace2012, Brown2019} and has been further advanced by contributions from Deutsch \cite{Deutsch1985, Deutsch1999, Deutsch2000, Kuypers2021}.

A central concept in Everettian quantum theory is the \textit{branching tree}, which represents the unitary evolution of the universe. Due to this branching structure, the theory has been popularly (and somewhat controversially among Everettians) known as the \textit{many-worlds interpretation} of quantum mechanics. Despite the popularity of the branching tree concept, the precise rules for dividing a system into \textit{relative states} have never been fully clarified. This ambiguity has led to a lack of consensus among Everettians regarding the branching process. 
While some argue that decoherence is necessary to define classical branches \cite{saunders2010}, recent developments indicate that \textit{foliations} are fully quantum \cite{Kuypers2021}. 
The most intriguing thought experiments occur in branching structures without decoherence, as full quantum coherence allows for the clearest comparison between different interpretations \cite{Nurgalieva2020}.

Recent progress by Kuypers and Deutsch involved formulating the relative state construction in the Heisenberg picture \cite{Kuypers2021}. 
Although both the Schrödinger and Heisenberg pictures are valid descriptions of quantum theory, the Heisenberg picture offers a precise description of when composite quantum systems are foliated. 
Additionally, it provides a clear definition of what constitutes an \textit{Everettian universe}, addressing a previously unresolved issue \cite{Vaidman2021}.
In the Schrödinger picture, relative states have been interpreted as leading to \textit{parallel universes}, which are seen as autonomously evolving components of the state vector. This view has led to the misconception that a sharp branching structure can always be constructed. However, reference \cite{Kuypers2021} demonstrated that, in general, a sharp branching structure does not exist. Insisting on foliating systems into parallel branches where it is not possible inevitably leads to paradoxes \cite{Gerhard2019}.

In this manuscript, we build upon reference \cite{Kuypers2021} to develop the relative descriptors for the Frauchiger-Renner thought experiment \cite{Frauchiger2018, Nurgalieva2020, Nurgalieva2022}, which remains a subject of ongoing debate across various interpretations \cite{Lazarovici2019, Polychronakos2024, delRio2024}. Despite the continued discussions, this thought experiment has revitalised the foundations of quantum mechanics, underscored the active role of physicists in probing quantum interpretations, and spurred further advancements in the field.

The experiment operates within a small Hilbert space and involves four \textit{quantum agents} that communicate with each other. As we will show, because the protocol is a decoherence-free extended Wigner's friend scenario, it serves as a prime example where a sharp branching tree is not feasible. This makes it an excellent exercise for working out the relative foliations for the thought experiment, while also addressing conceptual issues related to the foliation process in Everettian quantum theory.

The paper is organised as follows. In section \ref{sec:descriptors} we state the necessary elements from the Heisenberg picture of quantum computation relevant to the thought experiment.
In section \ref{sec:protocol} we review the concept of a quantum agent and the Frauchiger-Renner thought experiment's protocol in the form of a quantum circuit.  In section \ref{sec:relative} we work out the relative descriptors for the protocol, which allows us to construct and discuss the branching structure. In section \ref{sec:discussion}, we discuss what statement of the Frauchiger-Renner no-go theorem is negated. We conclude in section \ref{sec:conclusion}. 
\ref{sec:glossary} contains a glossary of technical terms intended to make the paper more accessible to a broader audience. We italicize the first occurrence of all terms in the main text that have corresponding entries in the glossary.

\section{Descriptors \label{sec:descriptors}}

The objective of our work is to determine the Heisenberg observables, or \textit{descriptors}, for the Frauchiger-Renner protocol for all qubits at all times; see figure \ref{fig:protocol}. 
In this section, we introduce the necessary descriptor preliminaries pertinent to the protocol. Except for the \textit{controlled-Hadamard}, details of this section's content may be consulted in references \cite{Deutsch2000,Horsman2007,Horsman2007n,Kuypers2021,Bedard2021abc,Bedard2021,bedard2023,cruz2024}. 
Readers familiar with descriptors may choose to skip this section.

A generic qubit \( Q \) is described by time-dependent Heisenberg observables, which are called descriptors and here denoted by \(\bm{q}_Q^{(t)}\). Qubit descriptors have three components, which at time \( t=0 \) coincide with the Schrödinger observables, and may be initialized as \(\bm{q}_Q^{(0)} = (\sigma_x, \sigma_y, \sigma_z)_Q\), where \( (\sigma_x, \sigma_y, \sigma_z) \) are the Pauli matrices. 
Descriptor components \( q_{Qi}^{(t)} \) with  \(i = x, y, z\)  obey the \textit{Pauli algebra} at all times, such that
\begin{equation}
q_{Qi}^{(t)}q_{Qj}^{(t)}=\delta_{ij}I+i\sum_k \epsilon_{ijk}q_{Qk}^{(t)},\quad 
\left[\boldsymbol{q}_Q^{(t)},\boldsymbol{q}_{Q'}^{(t)}\right]=0,\quad (Q\neq Q'),
\label{eq:algebra}
\end{equation}
where \( I \) is the identity, and \( \epsilon_{ijk} \) is the Levi-Civita symbol. The algebra allows us to omit one of the descriptor components for brevity, since that component is easily recovered from the other two.

\subsection{Unitary gates}

Looking at the quantum gates present in figure \ref{fig:protocol},
we must know how the rotation, the Hadamard, the controlled-not, and the controlled-Hadamard evolve the descriptor components. 
Given a generic quantum gate $G$, its effect on a descriptor component is given by the unitary Heisenberg evolution
\begin{equation}
G:\quad q_{Qi}^{(t+1)}=U_G^\dag\left[\boldsymbol{q}_{Q_1}^{(t)},\dots,\boldsymbol{q}_{Q_N}^{(t)} \right ]q_{Qi}^{(t)}\,U_G\left[\boldsymbol{q}_{Q_1}^{(t)},\dots,\boldsymbol{q}_{Q_N}^{(t)} \right ],
\label{eq:functional}
\end{equation}
where $\left\{\boldsymbol{q}_{Q_1}^{(t)},\dots,\boldsymbol{q}_{Q_N}^{(t)} \right\}$ is the set of all descriptors in an $N$ qubit computational network, and $U_G\left[\boldsymbol{q}_{Q_1}^{(t)},\dots,\boldsymbol{q}_{Q_N}^{(t)} \right ]$ is called the functional representation of $G$. 

We begin with the single qubit gates.
For the rotation about the $y$ axis by an angle $\varphi$ on a qubit $Q$, the functional representation reads
\begin{equation}
U_{R_y(\varphi)}\left[\boldsymbol{q}_Q^{(t)} \right ]=I\cos\left(\frac{\varphi}{2} \right )-iq_{Qy}^{(t)}\sin\left(\frac{\varphi}{2} \right ).
\end{equation}
Then, we may use equation (\ref{eq:functional}) together with the algebra in equation (\ref{eq:algebra}) to work out the effect of the rotation on the descriptor components:
\begin{equation}
R_{y}\left(\varphi\right):\quad \boldsymbol{q}_Q^{(t+1)}=\left(q_{Qx}^{(t)}\cos\varphi+q_{Qz}^{(t)}\sin\varphi,q_{Qy}^{(t)},q_{Qz}^{(t)}\cos\varphi-q_{Qx}^{(t)}\sin\varphi \right ).
\label{eq:rotation}
\end{equation}
Similarly, the functional representation for the Hadamard on $Q$ is
\begin{equation}
U_{H}\left[\boldsymbol{q}_Q^{(t)} \right ]=\frac{q_{Qx}^{(t)}+q_{Qz}^{(t)}}{\sqrt{2}},
\label{eq:Hadamard}
\end{equation}
and the Hadamard evolves $Q$ as:
\begin{equation}
H:\quad \boldsymbol{q}_Q^{(t+1)}=\left(q_{Qz}^{(t)},-q_{Qy}^{(t)},q_{Qx}^{(t)} \right ).
\end{equation}

We now turn to the evolution of the controlled operations, which act on a control qubit $C$ and a target $T$. The controlled-not gate evolves those qubits as:
\begin{eqnarray}
    \mathrm{controlled-not}:\quad  &\boldsymbol{q}_C^{(t+1)} = & \left(q_{C x}^{(t)} \underline{q_{T x}^{(t)}}, q_{C y}^{(t)} \underline{q_{T x}^{(t)}}, q_{C z}^{(t)} \right); \nonumber \\
    &  \boldsymbol{q}_T^{(t+1)}= & \left(q_{T x}^{(t)}, q_{T y}^{(t)} \underline{q_{C z}^{(t)}}, q_{T z}^{(t)} \underline{q_{C z}^{(t)}} \right),
\label{eq:cnot}
\end{eqnarray}
where the underlined terms highlight the parts that were copied from the other qubit during the interaction. 
The controlled-Hadamard has not been worked out in the literature in the Heisenberg picture, but it is straightforward to do so. We obtain:
\footnotesize
\begin{eqnarray}
& \mathrm{controlled-}H: \nonumber \\
& q_C^{(t+1)} = \left(q_{Cx}^{(t)} U_{H} \left[\bm{q}_T^{(t)} \right ], q_{Cy}^{(t)} U_{H} \left[\bm{q}_T^{(t)} \right ], q_{Cz}^{(t)} \right); \nonumber \\
& q_T^{(t+1)} = \left(q_{T x}^{(t)} P_1 \left[q_{Cz}^{(t)} \right ] + q_{T z}^{(t)} P_{-1} \left[q_{Cz}^{(t)} \right ], q_{T y}^{(t)} q_{C z}^{(t)}, q_{T z}^{(t)} P_1 \left[q_{Cz}^{(t)} \right ] + q_{T x}^{(t)} P_{-1} \left[q_{Cz}^{(t)} \right ] \right), 
\label{eq:ch}
\end{eqnarray}
\normalsize
where
\begin{equation}
P_{\pm 1}\left[q_{Cz}^{(t)} \right ]=\frac{1}{2}\left(I\pm q_{Cz}^{(t)} \right ),
\end{equation}
are the projectors in the Heisenberg picture and form a \textit{projection-valued measure}.

\subsection{Entanglement and foliations}

In the Heisenberg picture, a composite system may only be foliated into \textit{parallel universes} if it is entangled. Entanglement is established through local interactions. Thus, what are known in the Schrödinger picture as parallel universes only spread through local interactions and have finite spatial extent. The Heisenberg picture shows that the foliations are better thought of as local bubbles that display Everettian multiplicity \cite{Kuypers2021}.
Two qubits $Q$ and $Q'$ are said to be entangled at time $t$, if there exists a pair of descriptor components such that \cite{Horsman2007,Horsman2007n,Kuypers2021}
\begin{equation}
\left\langle q_{Qi}^{(t)}q_{Q'j}^{(t)}\right\rangle \neq 
\left\langle q_{Qi}^{(t)}\right\rangle\left\langle q_{Q'j}^{(t)}\right\rangle.
\label{eq:entanglement}
\end{equation}
The expectation values $\langle \ldots \rangle$ are taken with respect to the Heisenberg state, which, without loss of generality, is fixed to $\ket{\boldsymbol{0}}$.

A two-party composite system, such as a control and target qubit, may be sharply foliated into relative states at a time $t$ if it satisfies two conditions. 
First, the expectation values of the projectors are non-zero:
\begin{equation}
\left\langle P_{\pm 1}\left[q_{Cz}^{(t)}\right]\right\rangle\neq 0\quad\mbox{and}\quad \left\langle P_{\pm 1}\left[q_{Tz}^{(t)}\right]\right\rangle\neq 0.
\end{equation}
Second, 
the product of the $z$ components of the control and target is sharp, that is
\begin{equation}
\left\langle q_{Cz}^{(t)}q_{Tz}^{(t)}\right\rangle =1.
\end{equation}

With these conditions, one then may define foliated target $(T_1,T_{-1})$ and control $(C_1,C_{-1})$ qubit instances described by their corresponding relative descriptors
\begin{eqnarray}
& T_{\pm 1}:\quad \bm{q}_{T,\pm 1}^{(t)} := \bm{q}_{T}^{(t)} P_{\pm 1} \left[q_{Cz}^{(t)}\right]; \nonumber \\
& C_{\pm 1}:\quad \bm{q}_{C,\pm 1}^{(t)} := \bm{q}_{C}^{(t)} P_{\pm 1} \left[q_{Tz}^{(t)}\right],
\end{eqnarray}
which define the foliations $C_1/T_1$ and $C_{-1}/T_{-1}$. 
For each foliation, one then defines a corresponding conditional expectation value. For instance, the conditional expectation value for the target, provided that the $z$ component of the control was measured to be $\pm 1$ is
\begin{equation}
\left\langle q_{Ti}^{(t)}\right\rangle_{T,\pm 1}:=\frac{\left\langle q_{Ti,\pm 1}^{(t)}\right\rangle }{\left\langle P_{\pm 1}\left[q_{Cz}^{(t)} \right ]\right\rangle }. 
\label{eq:conditional}
\end{equation}
The relative descriptors obey a Pauli algebra of reduced dimensionality. For instance, for the target on the $C_1/T_1$ foliation, this reads \cite{Kuypers2021}: 
\begin{equation} q_{Ti,1}^{(t)}q_{Tj,1}^{(t)}=\delta_{ij}P_{ 1} \left[q_{Cz}^{(t)}\right]+i\sum_k \epsilon_{ijk},q_{Tk,1}^{(t)}, \end{equation}
where the only difference with respect to equation (\ref{eq:algebra}) is a redefined identity.

\section{Quantum agents and the protocol \label{sec:protocol}}

In this section, 
we introduce the concept of quantum agents, and outline the Frauchiger-Renner protocol. We assume the reader is familiar with the thought experiment, as it has been extensively discussed in the literature, not only by the original authors \cite{Frauchiger2018, Nurgalieva2020, Nurgalieva2022}, but also by others \cite{Lazarovici2019, Bub2021, Waaijer2021}.

\subsection{Quantum agents}

A \textit{quantum agent} is a quantum system that can observe another quantum system. 
The concept originates from Wigner's friend thought experiment \cite{Wigner1967} where a quantum agent (typically represented by Wigner's friend) refers to a quantum system capable of observing another quantum system through a unitary interaction, while simultaneously being subject to observation by an external observer (such as Wigner). 
This underscores the fundamental difference between a quantum agent's process of observation and memory formation \cite{Baumann2024}, and decoherent projective measurements. Notably, even though the interaction is unitary, within the framework of Everettian quantum theory, the agent is still able to establish an internal memory \cite{Deutsch1985,Kuypers2021}. 
Therefore, in this paper, when we say that an agent performs a measurement, we mean that the agent's corresponding qubit undergoes a unitary interaction via a controlled-not gate.

In extended Wigner's friend scenarios, such as the Frauchiger-Renner protocol, it is crucial to recognize that the observers can be modelled as qubits, rather than as classical observers. In realistic experimental settings, observation typically involves a pointer basis that is selected through decoherence \cite{Schlosshauer2007}. While this aspect lies beyond the scope of the present thought experiment, readers interested in a detailed example may check reference \cite{bedard2023}.

One might wonder whether there is a basis problem in the decoherence-free setup, but this is not the case. The reason is that entanglement determines the foliations, and entanglement cannot be undone by a change of basis \cite{Griffiths2018}. While the entire protocol could be arbitrarily rotated, the relative states would remain unaffected.

\subsection{The Frauchiger-Renner protocol}

\begin{figure}
\centering
\includegraphics[width=0.75\textwidth]{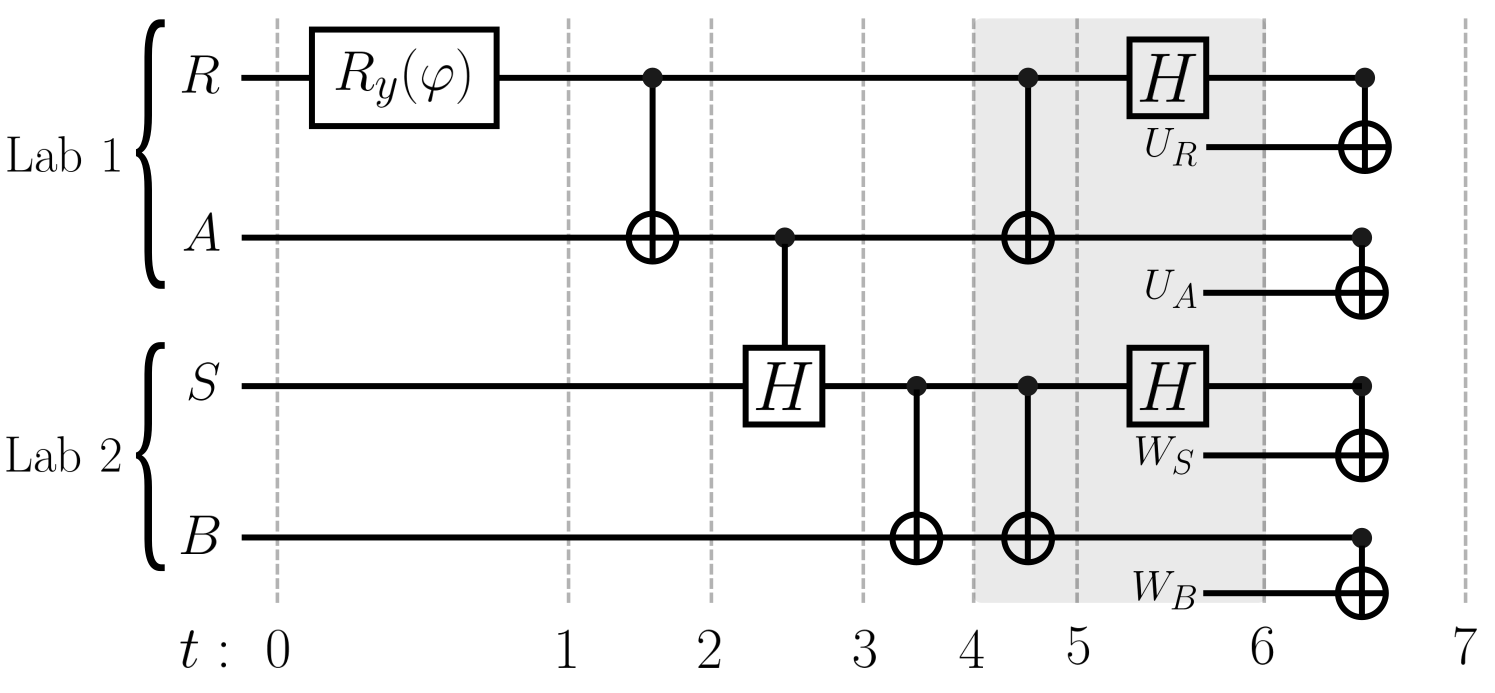}
\caption{
The Frauchiger-Renner protocol. 
The grey region prepares the Bell-basis. The qubit legs for Ursula and Wigner are only displayed at the end for clarity.
}\label{fig:protocol}
\end{figure}

The protocol may be represented by a quantum circuit; see figure \ref{fig:protocol}. The system is modelled by eight qubits $\{R,A,S,B,U_R,U_A,W_S,W_B\}$, where the last four only participate in the last step.
Two labs are under the full quantum control of external agents. The first lab contains qubits $R$ and $A$, where $A$ represents Alice's memory, and $R$ is a qubit in her lab.
The second lab has qubits $S$ and $B$, where $B$ represents Bob's memory, and $S$ is his qubit. Two external agents, Ursula $U$ and Wigner $W$ have full quantum control over Alice's lab $RA$ and Bob's lab $SB$, respectively. Ursula, to be able to measure both $R$ and $A$, must consist of at least two qubits herself. We call Ursula's qubit that interacts with $R$ by $U_R$, and her other qubit that interacts with $A$ by $U_A$. Wigner has an analogous structure with $W_S$ and $W_B$. 
A Qiskit implementation of the circuit in figure \ref{fig:protocol} may be found in reference \cite{Violaris_yt_23}. 

We now describe the protocol of figure \ref{fig:protocol}. All qubits initialise in the state $\ket{0}$. In the Heisenberg picture, we may therefore represent the fixed Heisenberg state of the eight qubit network as $\ket{\boldsymbol{0}}$. 
In the time interval $t\in (0,1)$, Alice's qubit $R$ evolves through a rotation about the $y$ axis $R_y(\varphi)$ by and angle $\varphi=2\arcsin(\sqrt{2/3})$. Next, at $t\in (1,2)$, Alice's unitary measurement on $R$ is modelled by a controlled-not gate, through which Alice records the eigenvalue $+1$ or $-1$ in her memory. Depending on the outcome, Alice prepares Bob's qubit $S$. If Alice's memory is $+1$, she leaves $S$ intact. But if Alice's memory is $-1$, then she operates a Hadamard gate $H$ on $S$. 
In the circuit, this is achieved by a controlled-Hadamard gate at $t\in (2,3)$, with $A$ as the control, and $S$ as the target. 
Figure \ref{fig:protocol} shows that the controlled-Hadamard is the only interaction that connects the two labs. 
At $t\in (3,4)$, Bob measures his qubit $S$ by means of a controlled-not gate. 
Between $t\in (4,7)$, Ursula and Wigner use the Bell-basis to measure Alice's and Bob's lab, respectively. 
The Bell-basis measurement is achieved by a controlled-not followed by a Hadamard \cite{Nielsen2010}, which is indicated by the grey region.
The implementation of this circuit shows that $U_A$ and $W_B$ always record the $+1$ eigenvalue, whereas $U_R$ and $W_S$ record both eigenvalues $\pm 1$ upon repetition of the protocol.

\section{Relative descriptors for the protocol \label{sec:relative}}

We now derive the descriptors for all stages of the protocol illustrated in figure \ref{fig:protocol}. 
For each time step, we then examine whether the emerging foliations are sharp or non-sharp.
This will allow us to construct the Everettian branching tree, which, in contrast to the fully distinct branches that arise in the presence of decoherence, will include both sharp and non-sharp regions.
The algebra of the descriptors in equation (\ref{eq:algebra}) allows us to omit the $y$ component of the descriptors for brevity, which we may easily recover when necessary.

The qubit $R$ starts out described by $\boldsymbol{q}_R^{(0)}=(\sigma_x,\sigma_z)_R$, and so do all other seven qubits. 
In the time interval $t\in (0,1)$, the rotation $R_y(\varphi)$ evolves $R$ to
\begin{equation}
\boldsymbol{q}_R^{(1)}=\left(\sigma_xc+\sigma_zs,\sigma_zc-\sigma_xs \right )_R,
\end{equation}
where $c=-1/3$ and $s=\sqrt{8}/3$, which satisfy $c^2+s^2=1$. The expectation value $\left\langle q_{Rz}^{(1)}\right\rangle =c$ shows that $R$ lost its sharpness. 

During $t\in (1,2)$, $A$ and $R$ interact through a controlled-not gate, with $R$ as the control, and $A$ as the target. Using equation (\ref{eq:cnot}), we find
\begin{eqnarray}
& \bm{q}_R^{(2)} = \left[\left(\sigma_x^R c + \sigma_z^R s \right) \sigma_x^A, \sigma_z^R c - \sigma_x^R s \right]; \nonumber \\
& \bm{q}_A^{(2)} = \left[\sigma_x^A, \sigma_z^A \left(\sigma_z^R c - \sigma_x^R s \right) \right].
\end{eqnarray}
With equation (\ref{eq:entanglement}) we may verify that $R$ and $A$ are now entangled. Moreover, given that the projectors have non-zero expectation value and that the product of the $z$ components is sharp, the composite system $RA$ at $t=2$ admits sharp foliations with relative descriptor components given by
\begin{equation}
q_{Az,\pm 1}^{(2)}=q_{Az}^{(2)}P_{\pm 1}\left[q_{Rz}^{(2)} \right ]=\pm\sigma_z^AP_{\pm 1}\left[q_{Rz}^{(2)} \right ].
\end{equation}
It is sufficient to write the relative descriptor of the $z$ component of the target. Using equation (\ref{eq:conditional}) we see that the conditional expectation values
$\left\langle q_{Az}^{(2)}\right\rangle_{A,\pm 1}=\pm 1$ are sharp. The analysis determines the two sharp foliations $R_{\pm 1}/A_{\pm 1}$; see figure \ref{fig:tree}. 

Based on these sharp outcomes, $A$ prepares the qubit $S$. The interaction between them is a controlled-Hadamard between $t\in (2,3)$, with $S$ as the target. Using equation (\ref{eq:ch}), we find 
\begin{eqnarray}
& \bm{q}_A^{(3)} = \left(\sigma_x^A U_H \left[\bm{q}_S^{(2)} \right ], q_{Az}^{(2)} \right); \nonumber \\
& \bm{q}_S^{(3)} = \left(\sigma_x^S P_1 \left[q_{Az}^{(2)} \right ] + \sigma_z^S P_{-1} \left[q_{Az}^{(2)} \right ], \sigma_z^S P_1 \left[q_{Az}^{(2)} \right ] + \sigma_x^S P_{-1} \left[q_{Az}^{(2)} \right ] \right).
\end{eqnarray}
We may work out the descriptor components in terms of the initial Pauli matrices. However, because the algebra of the descriptors is valid at all times, it is not always advantageous nor necessary to do so. 
$A$ and $S$ are entangled at $t=3$, because 
\begin{equation}
    \left\langle q_{Az}^{(3)} q_{Sz}^{(3)}\right\rangle \neq \left\langle q_{Az}^{(3)} \right\rangle  \left\langle  q_{Sz}^{(3)}\right\rangle .
\end{equation}
Nonetheless, the product $\left\langle q_{Az}^{(3)} q_{Sz}^{(3)}\right\rangle =1/3$, which does not yield sharp foliations between $A$ and $S$. Not all entangled systems foliate parallelly, which is possible but much harder to see in the Schrödinger picture. 
Note that at $t=3$, $R_{\pm 1}/A_{\pm 1}$ remains sharp. 
It is the $A/S$ part that forms a \textit{local interference bubble}. 

It is instructive to compare the Heisenberg and Schrödinger pictures at $t=3$. The Schrödinger state vector for $R$, $A$ and $S$ at $t=3$ is \cite{Nurgalieva2020}
\begin{equation}
    \frac{1}{\sqrt{3}}|1\rangle_R|1\rangle_A|1\rangle_S+\sqrt{\frac{2}{3}}|-1\rangle_R|-1\rangle_A
\,\frac{1}{\sqrt{2}}\left(|1\rangle+|-1\rangle \right )_S.
\label{eq:vector}
\end{equation}
In contrast to the Heisenberg picture, in equation (\ref{eq:vector}), the local state specification of each subsystem $R$, $A$, and $S$ is unavailable. This makes it harder to see that, although the system is entangled, the $R/A$ relative states are sharp, but the $A/S$ relative states are non-sharp; see figure \ref{fig:tree} at $t=3$.
Another advantage of the Heisenberg picture is that, notwithstanding entanglement, the global density matrix can be reconstructed from the descriptors of the subsystems, whereas in the Schrödinger picture, reconstructing it would require the global state vector \cite{bedard2023}.

At $t\in (3,4)$, $B$ measures his qubit $S$ through a controlled-not operation, with $B$ as the target. The descriptors at time $t=4$ are
\begin{equation} \boldsymbol{q}_S^{(4)}=\left(q_{Sx}^{(3)}\sigma_x^B, q_{Sz}^{(3)}\right ); \qquad
\boldsymbol{q}_B^{(4)}=\left(\sigma_x^B, \sigma_z^B q_{Sz}^{(3)}\right ).
\end{equation}
This allows for the relative instances
\begin{equation}
q_{Bz,\pm 1}^{(4)}=\pm\sigma_z^B P_{\pm 1}\left[q_{Sz}^{(4)} \right ],
\end{equation}
defining the sharp foliations $S_{\pm 1}/B_{\pm 1}$. 

During $t\in (4,6)$, Ursula and Wigner perform a \textit{Bell-measurement} on the $RA$ and $SB$ labs, respectively. The first stage includes a controlled-not between $R$ and $A$, and between $S$ and $B$, followed by a Hadamard on $R$ and $S$. This evolves the involved qubits to 
\begin{eqnarray}
& \bm{q}_R^{(6)} = \left(q_{Rz}^{(2)}, q_{Rx}^{(2)} q_{Ax}^{(3)} \right); \quad 
\bm{q}_A^{(6)} = \left(q_{Ax}^{(3)}, q_{Az}^{(2)} q_{Rz}^{(2)} \right); \nonumber \\
& \bm{q}_S^{(6)} = \left(q_{Sz}^{(3)}, q_{Sx}^{(4)} \sigma_x^B \right); \quad 
\bm{q}_B^{(6)} = \left(\sigma_x^B, q_{Bz}^{(4)} q_{Sz}^{(3)} \right).
\end{eqnarray}
This stage (before $U$ and $W$ interact) diffuses the $R_{\pm 1}/A_{\pm 1}$ and $S_{\pm 1}/B_{\pm 1}$ foliations that immediately before were sharp. 

In the last step at $t\in (6,7)$, Ursula and Wigner interact with their respective labs. These interactions are controlled-not gates between $R$ and $U_R$, $A$ and $U_A$, $S$ and $W_S$, and $B$ and $W_B$. 
The resulting descriptors for all eight qubits are
\begin{eqnarray}
& \bm{q}_R^{(7)} = \left(q_{Rz}^{(2)} \sigma_x^{U_R}, q_{Rx}^{(2)} q_{Ax}^{(3)} \right); \quad 
\bm{q}_{U_R}^{(7)} = \left(\sigma_x^{U_R}, \sigma_z^{U_R} q_{Rx}^{(2)} q_{Ax}^{(3)} \right); \nonumber \\
& \bm{q}_A^{(7)} = \left(q_{Ax}^{(3)} \sigma_x^{U_A}, q_{Az}^{(2)} q_{Rz}^{(2)} \right); \quad 
\bm{q}_{U_A}^{(7)} = \left(\sigma_x^{U_A}, \sigma_z^{U_A} q_{Az}^{(2)} q_{Rz}^{(2)} \right); \nonumber \\
& \bm{q}_S^{(7)} = \left(q_{Sz}^{(3)} \sigma_x^{W_S}, q_{Sx}^{(4)} \sigma_x^B \right); \quad 
\bm{q}_{W_S}^{(7)} = \left(\sigma_x^{W_S}, \sigma_z^{W_S} q_{Sx}^{(4)} \sigma_x^B \right); \nonumber \\
& \bm{q}_B^{(7)} = \left(\sigma_x^B \sigma_x^{W_B}, q_{Bz}^{(4)} q_{Sz}^{(3)} \right); \quad 
\bm{q}_{W_B}^{(7)} = \left(\sigma_x^{W_B}, \sigma_z^{W_B} q_{Bz}^{(4)} q_{Sz}^{(3)} \right).
\end{eqnarray}
All qubits at $t=7$ admit sharp relative descriptors, which for the final targets are
\begin{eqnarray}
& q_{U_R,z,\pm 1}^{(7)} = \pm \sigma_z^{U_R} P_{\pm 1} \left[q_{Rz}^{(7)} \right]; \quad 
q_{U_A,z,1}^{(7)} = \sigma_z^{U_A} P_{1} \left[q_{Az}^{(7)} \right]; \nonumber \\
& q_{W_S,z,\pm 1}^{(7)} = \pm \sigma_z^{W_S} P_{\pm 1} \left[q_{Sz}^{(7)} \right]; \quad 
q_{W_B,z,1}^{(7)} = \sigma_z^{W_B} P_{1} \left[q_{Bz}^{(7)} \right].
\end{eqnarray}
Note that $U_A$ and $W_B$ are guaranteed to observe the $+1$ eigenvalue. $U_R$ and $W_B$ may distinguish between Bell states. 
The resulting foliations are $R_{\pm 1}/{U_R}_{\pm 1}/A_1/{U_A}_1$ and $S_{\pm 1}/{W_S}_{\pm 1}/B_1/{W_B}_1$.

\begin{table}
\centering
\begin{tabular}{ccccc}
\hline
Time                 & Interaction          & Gate                 & Foliations           & Projections          \\ \hline
$(0,1)$              & -                    & Rotation on $R$      & -                    & -                    \\
$(1,2)$              & $R,A$                & Controlled-not       & Sharp                & $1/3,2/3$            \\
$(2,3)$              & $A,S$                & Controlled-$H$       & Non-sharp            & $1/3,2/3$            \\
$(3,4)$              & $S,B$                & Controlled-not       & Sharp                & $1/3,2/3$            \\
\multicolumn{1}{l}{} & \multicolumn{1}{l}{} & \multicolumn{1}{l}{} & \multicolumn{1}{l}{} & \multicolumn{1}{l}{} \\
$(4,5)$              & $R,A$                & Controlled-not       & Non-sharp            & $1/3,2/3$            \\
$(4,5)$              & $S,B$                & Controlled-not       & Non-sharp            & $1/3,2/3$            \\
$(5,6)$              & $R,A$                & Hadamard on $R$      & Non-sharp            & $5/6,1/6$            \\
$(5,6)$              & $S,B$                & Hadamard on $S$      & Non-sharp            & $5/6,1/6$            \\
\multicolumn{1}{l}{} & \multicolumn{1}{l}{} & \multicolumn{1}{l}{} & \multicolumn{1}{l}{} & \multicolumn{1}{l}{} \\
$(6,7)$              & $R,U_R$              & Controlled-not       & Sharp                & $5/6,1/6$            \\
$(6,7)$              & $A,U_A$              & Controlled-not       & Sharp                & $1,0$                \\
$(6,7)$              & $S,W_S$              & Controlled-not       & Sharp                & $5/6,1/6$            \\
$(6,7)$              & $B,W_B$              & Controlled-not       & Sharp                & $1,0$                \\ \hline
\end{tabular}
\caption{Summary of the time intervals $(t_1,t_2)$, the parties involved $(X,Y)$, the operation on the parties, the sharpness of the resulting foliations, and the values of the expectation values of the projections $\left\langle P_1\left[q_{Xz}^{(t_2)} \right ] \right\rangle $ and $\left\langle P_{-1}\left[q_{Xz}^{(t_2)} \right ] \right\rangle$.}
\label{tab:summary}
\end{table}

\subsection{Everettian foliations}

We summarise this section's results in table \ref{tab:summary}. In the last column we also state all the expectation values for the projectors for all interactions at each time step. These results are in agreement with the projections known from the Schrödinger picture \cite{Nurgalieva2020}.  

\begin{figure}
\centering
\includegraphics[width=0.9\textwidth]{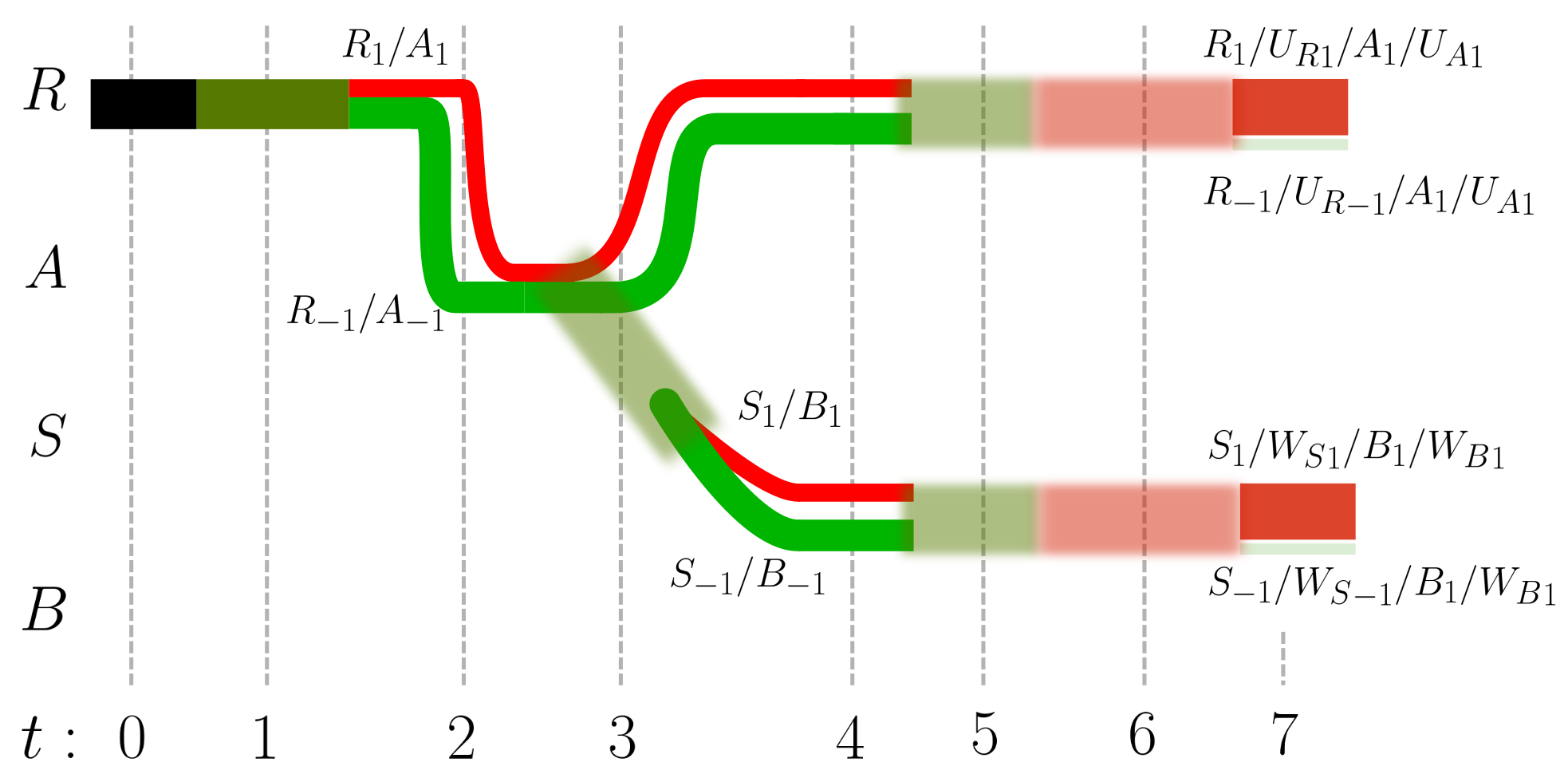}
\caption{
Everettian branching tree based on the relative descriptor construction for the Frauchiger-Renner protocol.
The branch thicknesses correspond to the projection values listed in Table \ref{tab:summary}. For example, the $R_1/A_1$ branch (red) has a thickness proportional to $1/3$, while the $R_{-1}/A_{-1}$ branch (green) has a thickness of $2/3$.
In non-sharp regions, the colours are blurred to represent the projection weights of the associated incoming or outgoing sharp branches. For instance, the non-sharp region associated with the outgoing branches $S_{\pm 1}/B_{\pm 1}$ is a blend of $2/3$ green and $1/3$ red, resulting in a pale green hue. 
}\label{fig:tree}
\end{figure}

Based on table \ref{tab:summary}, 
in figure \ref{fig:tree} we draw the Everettian branching tree for the Frauchiger-Renner protocol. 
In contrast to the typical imagery associated to Everettian branching trees, here the branching tree has non-sharp parts and all foliations are spatially local.
In the interval $t\in (0,1)$ the Hadamard changes the expectation value of the $R$ qubit, which is indicated by a colour change (black to dark-green). 
The black trunk of the tree contains instances of all eight qubits, but we emphasise the specific subsystems that are being foliated. 
Examining all descriptors at $t=2$ verifies that when $R$ and $A$ foliate, all other systems remain unaffected. 
The first sharp foliation occurs at $t\in (1,2)$, when the controlled-not originates $R_{\pm 1}/A_{\pm 1}$. 
The projection weights of these foliations may be consulted in table \ref{tab:summary}.
These foliations remain unaltered until $t\in (4,5)$, when another controlled-not at the same physical location diffuses $R_{\pm 1}/A_{\pm 1}$. 
Between $t\in(2,3)$, the controlled-Hadamard among $A$ and $S$ leads to the first local interference bubble, which we indicate by the blurry part with the colour reflecting the projection weights from the sharp foliations entering and emerging from the region. 
Whatever enters the non-sharp (blurry) region contributes through interference to the outcomes of the foliations that emerge from it at a later time.
Note that until $t\in (3,4)$, the same instance of $S$ and $B$ were carried through all branches of the tree until they finally foliated into $S_{\pm 1}/B_{\pm 1}$ due to a controlled-not at their physical location. 
At $t\in (4,5)$ both the $R_{\pm 1}/A_{\pm 1}$ and $S_{\pm 1}/B_{\pm 1}$ foliations diffuse through controlled-not gates. Between $t\in (5,6)$ the Hadamard on $R$ and $S$ changes the projections, which we again indicate with the colour transition. The final measurements by Ursula and Wigner lead to the final foliations $R_{\pm 1}/{U_R}_{\pm 1}/A_1/{U_A}_1$ and $S_{\pm 1}/{W_S}_{\pm 1}/B_1/{W_B}_1$.

\section{Discussion \label{sec:discussion}}

The Frauchiger-Renner thought experiment stirred the communities' interest, because they showed that at least one of three seemingly conservative statements must be false. We recall the statements here: (Q) Any agent may describe any system around them using the formalism of quantum theory and, based on this description, infer predictions (or retrodictions) of measurement outcomes using the quantum-mechanical Born rule; (C) An agent can use conclusions obtained by admitting the view of another agent; (S) A measurement carried out by an agent has a definitive outcome from the viewpoint of the measuring agent.
It is known that Everettian quantum theory rejects at least one of the rules \cite{Frauchiger2018}. In light of the tree in figure \ref{fig:tree}, we may comment on the statements in the present context. Here, (S) remains true. To evaluate (C),
the original argument uses Alice's relative outcome $R_{-1}/A_{-1}$ to infer that $B_1/{W_B}_1$. Yet, quantum theory admits both $B_{\pm 1}/{W_B}_{\pm 1}$. The reason is that both foliations $R_{\pm 1}/A_{\pm 1}$ contribute to $B_{\pm 1}/{W_B}_{\pm 1}$ through a non-sharp region. 
The same happens for the pilot-wave in Bohmian mechanics \cite{Lazarovici2019}.
The contradiction with quantum mechanics only arises if one insists having sharp foliations at all times, which is not possible in the Everettian theory. 
Wigner cannot adopt the viewpoint (C) of Alice/Bob, because they are using incompatible bases. Condition (Q) is emergent, but not fundamental in unitary quantum theory.

The results presented here rely on the assumption of a decoherence-free setup, enabling purely unitary evolution and providing a clear framework for analysing the Frauchiger-Renner thought experiment. While this idealized scenario highlights fundamental aspects of Everettian mechanics, it may limit the direct applicability of our findings to real-world systems, where environmental decoherence is inevitable. Addressing this limitation would involve technical challenges similar to those in quantum computing, where error correction encodes quantum information into a larger set of physical qubits. Analogously, the memory registers of quantum agents would require encoding with additional ancilla qubits to maintain coherence. Future research could explore the behaviour of quantum agents' memory registers under environmental decoherence. This motivates a research program for a quantum artificial intelligence programmed as a quantum agent.

\section{Conclusion \label{sec:conclusion}}

We worked out the relative descriptors for the Frauchiger-Renner thought experiment. 
In the context of Everettian quantum mechanics,
the exercise shows that in experiments where agents have full quantum control over other agents, the branching tree will have both sharp and non-sharp parts. Sharp relative foliations may be defined from entangled composite systems, but not every entangled system foliates sharply.

\section*{Data availability statement}
All data supporting the findings of this study are available within the paper.

\ack
The authors acknowledge the funding from CNPq, FAPERGS and BIC UFRGS.

\section*{Conflict of interest}
The authors declare no conflict of interest.

\appendix
\section{Glossary of Key Terms \label{sec:glossary}}

To reduce the paper's technical jargon and make it more accessible to a broader audience, we provide the following glossary.

\begin{itemize}

\item \textbf{Everettian quantum theory}: The interpretation of quantum mechanics that posits that Hilbert space maps to reality. 

\item \textbf{Decoherence-free setup}: A scenario where quantum coherence is preserved, and the effects of environmental decoherence are absent, allowing for purely unitary evolution.

\item \textbf{Relative state}: The state of one part of a composite quantum system that is described relative to the state of another part. 
    
\item \textbf{Foliation}: 
  A term borrowed from differential geometry and complex analysis, referring to the decomposition of a space into lower-dimensional subspaces (foliations), which provides a more manageable way to study its global properties. In quantum theory, foliations can be identified by examining the relative states of entangled subsystems.

\item \textbf{Branching tree}: A representation of the unitary evolution of a quantum system in Everettian mechanics, 
  showing how the universe foliates. 

  \item \textbf{Everettian universe}: A fully quantum, spatially local entity described in the many-worlds interpretation of quantum mechanics. It represents one of the quasi-parallel, dominantly autonomously evolving components of the universal wavefunction, emerging through measurement and entanglement processes. Everettian universes can exhibit quasi-classical behaviour under specific conditions, such as when certain observables are sharp and follow classical dynamics.

\item \textbf{Descriptor (Heisenberg observable)}: Time-dependent operators in the Heisenberg picture that describe the state of a quantum system. These evolve under the system's dynamics.
    
\item \textbf{Pauli algebra}: The algebra of Pauli matrices, fundamental in describing quantum mechanics of spin-$1/2$ particles and qubits.

\item \textbf{Controlled-Hadamard gate}: A quantum logic gate that applies a Hadamard operation to a target qubit conditional on the state of a control qubit. It is less commonly studied in the Heisenberg picture.

\item \textbf{Projection-valued measure}: A set of projection operators where the sum of all projection operators is the identity operator.

\item \textbf{Sharp vs. non-sharp foliations}: Sharp foliations result in distinct, well-defined branches. Non-sharp foliations arise when quantum interference prevents clear separation.

\item \textbf{Quantum agent}: A quantum system capable of making observations and storing the results as memory through unitary interactions, such as those modelled in Wigner's friend scenarios.

\item \textbf{Parallel universe}: A special type of foliation that occurs
when a composite system foliates into sharp instances without subsequent interference between them. In the many-worlds interpretation, such foliations are historically referred to as parallel universes. However, a quantum agent can access other foliations through non-sharp regions (interference bubbles).

\item \textbf{Local interference bubble}: A region in a branching structure where entanglement creates non-sharp foliations, indicating interference between branches.

\item \textbf{Bell-basis measurement}: A quantum measurement performed in the Bell basis, commonly used in entanglement protocols to distinguish maximally entangled states.
\end{itemize}

\newpage
\bibliographystyle{unsrt}
\bibliography{references} 
\end{document}